\documentstyle[aps,preprint,epsf]{revtex}

\draft

\begin{document}

\title{Explaining the Uneven Distribution of Numbers in Nature }

\author{L.Pietronero$^1$, E. Tosatti$^{2,3}$, V. Tosatti$^2$ 
and A. Vespignani$^3$}

\address{$^1$Dipartimento di Fisica and Unit\'a INFM dell'Universit\'a di
Roma ``La Sapienza'',\\
P.le A. Moro 2, I-00185 Roma, Italy\\
$^2$ SISSA/ISAS, Via Beirut 2-4, 34014 Trieste, Italy\\
$^3$ The Abdus Salam International Centre for Theoretical Physics (ICTP),
P.O.Box 586, 34100 Trieste, Italy}

\maketitle
\date{pttv.tex ~~~ \today ~~~ DRAFT}

\noindent
%{\bf 
Suppose you look at today's stock prices and bet on the value of 
the first digit. One could guess that a fair bet should correspond 
to the frequency of $1/9 = 11.11\%$ for each digit from $1$ to $9$. This 
is by no means the case, and one can easily observe a strong 
prevalence of the small values over the large ones. The first three 
integers $1$,$2$ and $3$ alone have globally a frequency of $60\%$ while the 
other six values $4$, $5$, $6$, $7$, $8$ and $9$ appear only in $40\%$ 
of the cases. This situation is actually much more general than 
the stock market and  it occurs in a variety of number catalogs 
related to natural phenomena. 
The first observation of this property traces back to S. Newcomb 
in 1881\cite{new} but a more precise account was given by 
F. Benford in 1938\cite{benf,ben}. 
He investigated $20$ tables of numbers ranging from the area of lakes and the 
length of rivers to the molecular weights of molecular compounds. 
In all cases he found the same behavior for which he guessed the 
probability distribution $P(n) = \log [(n+1)/n]$ where $n$ is the value 
of the first integer. Since then this observation has remained 
marginal and occasionally reported as a mathematical curiosity\cite{raimi}
with some computer science applications\cite{scha} and even tax-fraud 
detection\cite{nigrini}. 
In this note we illustrate these observations with the enlightening 
specific example of the stock market\cite{ley,hill}.
We also identify the general mechanism for the origin of 
this uneven distribution in the multiplicative nature 
of fluctuations in economics and in many natural phenomena. 
This provides a natural explanation for the ubiquitous presence
of the Benford's law in many different phenomena with the common
element that their fluctuations refer to a fraction of their values. 
This brings us close to the problem of the spontaneous origin of 
scale invariant properties in various phenomena which is a debated 
question at the frontier of different fields.
%}
Consider the values of the Athens, Madrid, Vienna and  Zurich stock 
markets of January 23, 1998. The stock prices $N$ are expressed 
in the local currencies. 
If the values of N were randomly distributed we would expect a uniform 
distribution for the value of the first digit $n = 1$, $2$,$.~.~.$, $9$. 
This would lead to $P(n) = 1/9 = 11.11\%$. In Fig.1a we can see instead 
that $P(1) = 30\%$ and then the frequency decreases continuously with $n$ 
until the minimum value $P(9) = 4.5\%$. This is the asymmetric distribution 
of first digits pointed out by F. Benford in 1938\cite{benf,ben} 
who investigated mostly tables of natural numbers. 
Here we can see that such a behavior applies to economic data as well. 
The prevalence of the number 1 in the first digit may suggest ad 
hoc explanations like the fact that there may be a tendency, 
in each country, to assign price values of the order of unity in 
local currency. This hypothesis can be easily checked by expressing 
the stock prices in a different currency. To this purpose, we 
repeat the analysis re-expressing the Madrid stocks in Swiss francs 
and the Zurich ones in Pesetas. Remarkably we observe that the Benford's 
distribution is independent on the units adopted. This fact may appear 
strange at first sight, but it is one of the most common example of 
scale invariance \cite{mand}.
This property should be found in the probability distribution of the 
original stock prices $P(N)$ and then it should reflects also in 
the distribution of their first digit $P(n)$. Changing the unit of 
measure of a price corresponds in  multiplying it by a factor $b$,
in our case the exchange rate between Swiss Francs and Pesetas. If we change
our units by a factor $b$, the values $N$ will become $N'=bN$ and the 
corresponding distribution should be identical to the original one, apart 
from a constant rescaling factor $A(b)$, that may depend upon $b$ but not 
on $N$. In mathematical terms scale invariance corresponds to the following 
functional relation 
\begin{equation}
P(N')=P(bN)=A(b)P(N).
\end{equation}
It is well known from statistical physics and critical phenomena that the 
general solution of this equation is any power law behavior with exponent 
$\alpha$
\begin{equation}
P(N')=N'^{-\alpha}=b^{-\alpha}N^{-\alpha}.
\end{equation}
For this type of distributions we can compute the probability of the 
first digit by observing that for each decade we have the same relative 
probability for the various integers, so we can write 
\begin{equation}
P(n)=\int_{n}^{n+1} N^{-\alpha}~dN=\frac{1}{1-\alpha}
[(n+1)^{(1-\alpha)}-(n)^{(1-\alpha)}],
\label{gen}
\end{equation}
for $\alpha\neq 1$. For $\alpha=1$, we have instead
\begin{equation}
P(n)=\int_{n}^{n+1} N^{-1}~dN=\int_{n}^{n+1}d(\log N) =\log(\frac{(n+1)}{n}),
\label{benny}
\end{equation}
that is precisely Benford's law as derived from the data analysis.
From Eq.~(\ref{benny}), we can see that the case $\alpha=1$ corresponds to 
a uniform distribution in logarithmic space as from Eq~(\ref{benny}). 
This means that if we consider a set of random numbers $R$ in normal 
space and then we look at the distribution $N=\exp(R)$, we have 
a uniform (random) distribution just in the logarithmic scale. 
This distribution 
is actually satisfying the Benford's law by construction\cite{nota1}. 

These observations pose therefore two questions:
i) The first is to understand why some data set naturally show 
scale invariant properties. ii) The second question concerns why, among the 
various scale invariant distributions corresponding to different $\alpha$ 
values, the $\alpha=1$ is the one actually realized in nature.
In the following we rationalize the two above questions in the 
finding of a general mathematical origin of the distribution 
$P(N)=1/N$ that can apply to a variety 
data set in several fields ranging from economics to physics and geology.

The most general mathematical property which applies to the statistics 
of a large variety of phenomena is the central limit theorem that governs the 
probability distribution corresponding to the sum of random numbers. From
a dynamical point of view we can consider the value of a variable 
$N$ which changes with 
time by the addition of a random variable $\xi$, yielding the Brownian
process
\begin{equation}
N(t+1)=\xi+N(t).
\end{equation} 
If the random variable $\xi$ is symmetrically distributed with
finite variance, the probability  distribution $P(N,t)$ to have a given value 
$N$ after $t$ additional  steps  will be  Gaussian with variance 
$\sigma\sim t^{1/2}$. In the infinite time limit the variance is diverging 
and the probability distribution will approach the uniform one.
This is indeed very far from the scale invariance properties we are looking 
for. On the other hand, it easy to realize that the Brownian dynamics does
not realistically apply to many stochastic dynamical phenomena.
Brownian dynamics is ruled by a noise term whose intensity 
is independent of the variable value $N$. Fluctuations are in this way
independent and related to some external dynamical parameters. 
Clearly, many systems do not follow such a 
dynamical description. For instance, it is intuitive to consider that a 
stock price has fluctuations which are relative to the price itself. 
In practice, each stock suffers of percent increments. Hence 
\begin{equation}
N(t+1)=\xi N(t),
\end{equation} 
where $\xi$ is again a stochastic variable that in this case must be 
positive definite. The nature of this process is completely
different from the usual Brownian motion. We can, however, relate the two
processes by a simple transformation. If we take the variable in logarithmic 
space we get:
\begin{equation}
\log N(t+1)=\log\xi+\log N(t)   ,
\end{equation} 
If we consider $\log\xi$ the new stochastic variable, we 
recover a Brownian dynamics in a logarithmic space; i.e. a random 
multiplicative process corresponds to a random additive process in logarithmic
space. This implies that for $t\to \infty$ the distribution $P(\log N)$ 
approaches a uniform distribution, and  by transforming  back to 
linear space we have 
\begin{equation}
\int P(\log N)d(\log N)=C\int\frac{1}{N}dN,
\end{equation}
where $C$ represents the normalization factor. This immediately gives 
$P(N)\sim N^{-1}$ as the distribution of  variable values $N$. 
Accordingly, the distribution of first digits $n$ will follow an 
ideal Benford law.

In order to illustrate numerically this result, we have considered a 
uniform distribution of numbers in the interval $(0,999)$. Clearly,
such a case corresponds to a uniform distribution for both the actual 
numbers and their first digits. We start to apply on the numbers 
of the starting uniform distribution a multiplicative dynamics as 
described previously. At each iteration step, Numbers are 
multiplied or divided by a factor $\Delta$ at random, following a bimodal 
distribution characterized by two 
delta function as $p(\xi)=\frac{1}{2}\delta(\xi-\Delta)+
\frac{1}{2}\delta(\xi-\Delta^{-1})$. In this way we have that in the 
logarithmic space the stochastic variable $\log\xi$ has zero mean and 
it is symmetrically distributed. Then we repeat the updating 
process of variables, which in logarithmic space corresponds
to the random addition of $\pm\log\Delta$. 
After enough multiplicative steps we can observe the resulting distribution
of number $N$. In Fig.2 it is reported the outcome of $100$ iteration
steps with $\Delta=1.5$. Both the first digit and the cumulative
numbers distribution evolved in the ideal Benford's law 
of Eq.~(\ref{benny}). 
This simple exercise shows that the numbers $N$ characterizing some 
physical quantities or objects, naturally will follow the Benford's law if
their time evolution is ruled by multiplicative fluctuations.
It is worth remarking that many  physical phenomena 
shows scale invariant behavior and are characterized by power 
law distributions. These features are due to cooperative effects, the onset 
of critical points and other nonlinear dynamical effects\cite{mand,bak}. 
Here, however, scale invariance manifests itself with non-trivial
power law exponents. Benford's law will apply to these phenomena
in the generalized form of Eq.~(\ref{gen}). In this sense, it is 
interesting to explore also connections with other well known scale invariant
features such as the Zipf's law\cite{zipf}. 

The fact that the Benford's law is naturally explained in terms of a 
dynamics governed by multiplicative fluctuations can provide new 
insights on many scale invariant natural phenomena. The understanding of 
the origin of scale invariance has been one of the fundamental tasks
of modern statistical physics. How system with many interacting 
degrees of freedom can spontaneously organize into critical or scale
invariant states  is a subject that is of upsurging interest to many 
researchers. Here we provide a simple dynamical picture of the generation 
of scale invariant distributions. 
A multiplicative dynamics might be sort of obvious in 
the stock pricing, but it is much less clear in the case of lake sizes or 
molecular weights of chemical compounds. 
\noindent{\bf Acknowledgments:}
We thank G.Mussardo for useful discussions and comments. 
L.P. thanks  the hospitality of ICTP where part of this work has 
been completed. L.P and A.V. acknowledge partial support from the 
TMR European Program ERBFMRXCT980183.

\begin{figure}[htb]
\centerline{
        \epsfbox{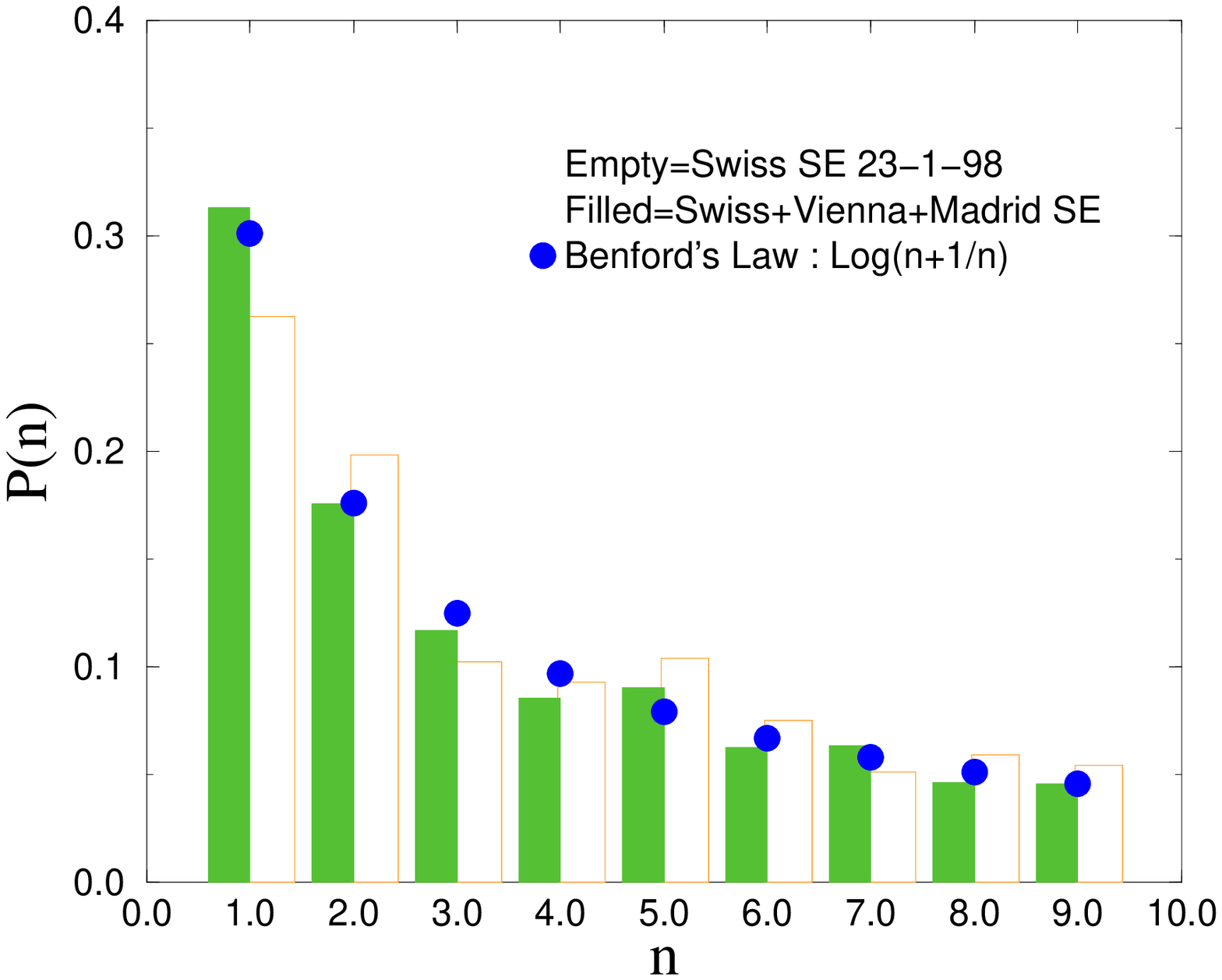}
        \vspace*{0.5cm}
%\newpage
%\centerline{
%        \epsfxsize=6.0cm
%        \epsfbox{.ps}
%        \vspace*{0.5cm}
%        }
%\centerline{
%        \epsfxsize=6.0cm
%        \epsfbox{fig1c.ps}
%        \vspace*{0.5cm}
        }
\caption{( Distribution $P(n)$ of the first digits of the stock prices $N$ 
of Zurich (expressed in Swiss francs). The distribution is strongly 
asymmetric and it is fairly reproduced by 
the Benford's law shown for comparison with circles. 
Deviations from the ideal Benford's law are due to statistical noise. Averaging
over three data sets (Madrid, Vienna and Zurich stock exchanges) we 
have that the distribution $P(n)$ shows smaller deviations.} 
\label{fig:1}
\end{figure}
\begin{figure}[htb]
\centerline{
        \epsfbox{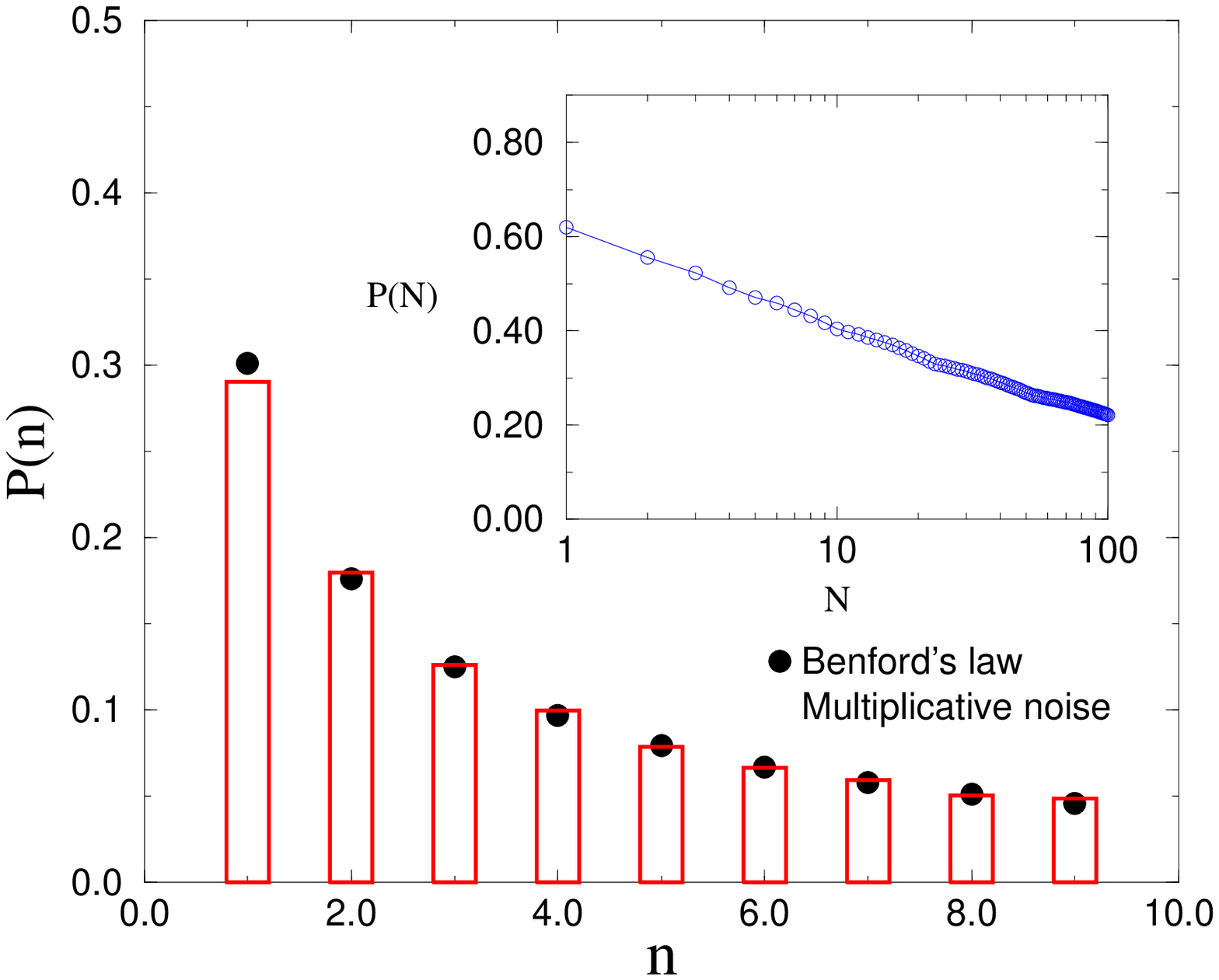}
        \vspace*{0.5cm}
       }
\caption{ Probability distribution of the first digit of a flat number
distribution after 100 iterative applications of  multiplicative noise 
(see text). The ideal Benford's law is strikingly satisfied. Small deviations 
disappear with more iterative noise applications. In the inset it is shown 
the cumulative distribution of all the numbers 
$P_c(N)=\int_N^{K^*} P(N')dN'$, where $K^*$ is the upper cut-off; i.e. the 
largest number generated. For a Benford-like distribution we have 
$P_c(N)=C(-\log N +\log K^*)$ where $C$ is the normalization factor. 
This behavior is nicely followed by the generated distribution 
as it appears from the linear plot on the semi-logarithmic scale. }
\label{fig:2}
\end{figure}


\begin{thebibliography}{99}
\bibitem{nota1}
	From this point of view it is easy to recognize that 
	also the geometric series obey 
	the Benford's law, as pointed out in the past\cite{ben}.
\end{thebibliography}
\end{document}